# Examination of Hydrogen Evolution Bubble Trapping in Ordered Porous 3D Printed Metal and Metal Oxide-Coated Microlattice Electrodes

Matthew Ferguson[a], Alex Lonergan[a], Christopher Kent[a], Dara Fitzpatrick[a], Colm O'Dwyer[a,b,c]*

[a] *School of Chemistry, University College Cork, Cork, T12 YN60, Ireland*
[b] *Sustainability Institute, University College Cork, Lee Road, Cork, T23 XE10, Ireland*
[c] *AMBER@CRANN, Trinity College Dublin, Dublin 2, Ireland*

**Abstract**

Determining the nature of surface roughness and electrode pore structure on $H_2$ bubble evolution rate and quantity, and bubble trapping under electrolytic conditions is important for quantifying useful gas production during total water splitting and hydrogen evolution reactions. Controlled electrode systems involving the design of geometry, surface area, and porosity provides options to understand trapped/redissolved gas bubble evolution and improve overall efficiency. In this study, we use vat polymerization (Vat-P) 3D-printing to create ordered microlattice electrode structures from metal and metal-oxide coated photopolymerized methacrylate-based resins. These micro-lattice structures are designed with various geometries to influence bubble traffic from gas nucleation and evolution during electrochemical HER processes. Using cyclic and linear sweep voltammetry, and chronopotentiometry, this work analyzes the response of metallized (NiO/Ni(OH)$_2$ and Au) microlattice HER electrodes as a function of geometric structure, to gauge influence of material activity, small scale surface roughness, and the larger substrate pore network on the traffic or larger bubbles formed during HER. This work also uses broadband acoustic resonance dissolution spectroscopy (BARDS) to quantify bubble evolution and reabsorption in the electrolyte during electrolysis. The results show that coated 3D printed electrodes are robust HER electrodes, allow efficient transport of small bubbles, but significant limitations are found for larger bubble transport through ordered porous microlattice shown through model simulations and experimental measurements.

# Introduction

In the pursuit of efficient green hydrogen production via electrochemical water splitting, the hydrogen evolution reaction (HER) depends not only on stable and active catalytic materials(*1*) but also on maximizing accessible surface area(*2, 3*). The development of sustainable electrode materials (*4*) is therefore essential to complement established hydrogen production technologies such as proton exchange membrane electrolyzers (*5*), solid oxide fuel cells, and alkaline water electrolyzers (*6*). The most common approach to high surface area in a volumetrically efficient manner is the use of porous metal and related materials (*7, 8*). Using porosity is not always straightforward, and bubble traffic is a paramount concern when optimizing collected hydrogen. For example, bubble trapping, control of bubble size and nucleation, coalescence and coverage of active surface area that lead to reduced overall activity and increased ohmic resistance, among many other negative effects of trapped bubble, have been studied in recent years (*9-11*). Addressing surface wetting energies, porous architectures, bubble size through current density control, surface energies, and other methods, have been tested to find a generally applicable explanation for minimizing these efficiency-reducing effects in HER systems (*12*).

Attracted by the freedom of design, ease of prototyping, and unprecedented ability to produce advanced architected components, 3D-printing has been increasingly integrated in recent years into the fields of materials science and electrochemistry (*13-15*), including batteries (*16-18*), supercapacitors (*19-22*) and water splitting applications (*23-25*). In HER systems, 3D printing enables the fabrication of electrodes with precisely defined porosity, offering a platform to investigate fundamental aspects of bubble transport within high–surface-area structures. Compared to bulk electrodes, these architectures allow the active catalytic material to be confined to thin coatings on lightweight scaffolds, substantially reducing material usage. Consequently, 3D-printed electrodes are now increasingly employed in electrochemical systems such as gas-flow reactors and electrolyzers (*26, 27*).

A variety of additively manufactured architectures including controlled-porosity current collectors, helical electrode platforms, octet-truss lattices, and cellular matrices have been explored for electrochemical applications (*23, 28-33*). Printed architectures have also been used in recent years as water splitting electrodes. In a study by Ambrosi et al. (*25*), a series of cage-like water-splitting electrodes were designed and 3D-printed by selective laser melting (SLM). Electrodeposition was used to coat the architected electrode surfaces in thin films of catalytically active materials, such as Ni, $IrO_2$, or Pt. This coating of the high-surface-area printed component greatly reduces the required catalytic material for a given area/volume. Other studies focus on how 3D-printed architectures can aid in the detachment and channelling of the gas bubbles produced from water electrolysis, counteracting the negative effects of bubble entrapment that may be encountered with porous material electrodes (*34-36*).

Using geometrically precise microlattice porous structures have several advantages over high porosity nanomaterials on planar surfaces (for example), including better mechanical properties, pore definition and interconnection, avoidance of closed cavities, and uniform pore distribution. These features enable systematic investigation of how surface area, geometry, pore shape, and effective density influence HER performance and gas evolution. (*7, 37, 38*). One important aspect of electrochemical HER is the controlled strength and homogeneous distribution of electric field current densities. Microlattice porous electrodes are well-suited to account for this, with the lattice features and truss connections being pre-defined in a known porous volume. We can assess the influence of this on bubble quantity and generation compared to randomly porous high surface area materials. Recent work has shown that 3D printing ordered porous structures can give useful information on gas flow in electrochemical flow reactors (*26*) and opportunities exist to examine the nature of hydrogen evolution using microlattice structures coated with material of different electrochemical activity to HER. Importantly, increasing porosity or surface area through intricate lattice design does not necessarily lead to improved performance in HER, oxygen evolution, or other gas-evolving reactions. Both ordered and random porous structures strongly influence gas evolution dynamics(*39*), and the rates of bubble nucleation, growth, coalescence, and detachment are critical factors in maximizing recoverable hydrogen.

Additive manufacturing enables the creation of larger, more open pore structures than conventional metal foams or wools, while simultaneously reducing electrode mass and catalyst loading. However, it remains unclear whether such large-pore architectures promote the formation of larger bubbles, or under what conditions bubble growth and transport can avoid coalescence within a high–surface-area network where pores may become restrictive once bubbles exceed their detachment diameter. In this work, we investigate how catalytic activity and ordered, interconnected pore architectures influence bubble transport and escape to the electrolyte–gas interface in HER systems. We demonstrate that bubble traffic critically affects electrochemical response, and that bubble trapping can negate the benefits of increased surface area in certain cases. Using a combination of simulation and operando acoustic resonance spectroscopy, we directly quantify evolved gas and elucidate the interplay between electrode architecture, bubble dynamics, and HER performance.

**Experimental**

*Printing and Processing of 3D Printed Microlattice Electrodes*

Vat polymerization 3D printing was implemented in this study to additively manufacture the water splitting lattice electrodes. The model of Vat-P printer used in this work is the FormLabs Form 2 (© Formlabs, Somerville, Massachusetts, USA), which uses a 405 nm laser (250 mW) as the light source, yielding a maximum printing resolution of 25 μm per layer. FormLabs Preform Clear V4 and High Temp V2 resins were chosen as the raw printing materials for these printed lattices.

Before printing, the lattices were designed using CAD software, with each of the lattices printed with dimensions of approximately 1 cm × 1 cm × 1 cm. The printing process (for 6 lattices) takes ~4 hours. Once printed, the lattices were gently removed from the building platform using a scraper and placed into an iso-propyl alcohol (IPA) bath, where they were washed thoroughly for at least 10 minutes to remove any residual liquid resin. Once washed, they were removed from the bath and dried. The lattices were then placed into a UV curing machine for a time and temperature specific to each resin type (15 minutes at 60 ºC for Clear V4 resin and 120 minutes at 80 ºC for High Temp V2 resin).

*Coating and Preparation of Printed Microlattice Electrodes*

The printed lattices were coated in a thin layer of electroactive material using metal sputtering. For this study electroactive materials of varying activity in neutral pH aqueous electrolyte (Ni and Au metal) were chosen as well understood species in water electrolysis. A Quorum Q150T S metal sputterer (Quorum Technologies, Laughton, East Sussex, UK), 99.98% purity Ni target (57 mm diameter, 0.1 mm thickness), and 99.99% purity Au target (57 mm diameter, 0.2 mm thickness) purchased from Ted Pella Inc. were used for the sputtering process. The cubic shaped lattices were coated in 50 nm of Ni or Au six times, each time with a different face pointed towards the metal target, ensuring all surfaces of the lattice truss features are entirely covered in metal. Following the metallization of the printed microlattices, the next step involved forming a waterproof electrical connection to a coated wire. Strips of coated wire (approximately 20–25 cm in length) were cut, with several mm of coating removed at each end. One end of the exposed wire was aligned with the planar base of the lattice, where several coatings of silver conductive lacquer were pasted to form a robust, electronically conductive contact between the wire and the lattice. Once the lacquer was left to dry for several hours, a layer of epoxy glue was pasted over the entire planar base of the lattice. The epoxy glue served to firmly attach the wire to the lattice base, waterproof the electrical connection, and ensure that only the lattice portion of the electrode is contributing to the electrochemical response during analyses. The sputtered and connected lattice electrodes can be seen in Figure 1(d).

*Electrochemical Setup and Analyses*

All electrochemical measurements were carried out in a standard 500 mL beaker, filled with 300 mL of pH neutral 1 M $K_2HPO_4/KH_2PO_4$ aqueous electrolyte. A three-electrode configuration was used for all electrochemical measurements, with the lattice electrodes acting as the working electrode (WE), a large planar Pt-sputtered Si-wafer acting as the counter electrode (CE) (*40*), and a Saturated Calomel electrode (SCE) acting as a reference (see Figure 1 (e)). Voltages were converted to values relative to RHE. A flexible electrode holder was used to keep consistent distances and configurations between electrodes across multiple tests. Cyclic voltammetry (CV), linear sweep voltammetry (LSV), and chronopotentiometry (CP) tests were carried out at room temperature using a BioLogic SP–150e potentiostat. The specific settings and parameters used for each electrochemical test can be found in the Supplementary Information.

*Materials Characterization*

Scanning electron microscopy (SEM) images of the lattice electrodes subjected to various electrochemical conditions were obtained using an FEI Quanta 650 FEG Scanning Electron Microscope, with and average chamber and gun pressure of $1.57 \times 10^{-4}$ mbar and $6.41 \times 10^{-9}$ mbar respectively. The samples were mounted onto aluminium SEM pin stubs (12 mm diameter, Agar Scientific, Essex, UK) and the chamber was pumped to a high vacuum. A working distance of 10 mm, a spot size of 3.5, and a beam voltage of 10 kV were used when acquiring images.

X-ray photoelectron spectroscopy (XPS) was carried out using a Kratos AXIS ULTRA spectrometer with a monochromatized Al Kα X-ray gun (1486.58 eV; 150 W, 10 mA, 15 kV). All samples were analyzed at temperatures of 20–30 ºC. For survey spectra, a pass energy of 160 eV, step of 1.0 eV, and dwell time of 50 ms was used. For narrow regions, a pass energy of 20 eV, step of 0.05 eV, and dwell time of 100 ms was used. Charge correction was carried out with respect to the C 1s photoemission at 284.8 eV. Core-level spectra were fitted with a Shirley-type background, and synthetic peaks were a mix of Gaussian-Lorentzian type. Relative sensitivity factors, containing Scofield cross-sections, were as follows: C 1s (1.0), O 1s (2.93), P 2p (1.19), Ni 2p (22.2), and K 2p (3.97).

Broadband acoustic resonance dissolution spectroscopy (BARDS) is an analytical technique that analyzes the change in resonant frequency when a solute is dissolved in a solvent. As the compressibility of a solution changes (due to dissolved gas, gas bubbles, etc.), so does the speed of sound in solution. The fundamental response and basis of BARDS can be found elsewhere (*41, 42*). Briefly, the evolution of hydrogen gas bubbles in the electrolyte has a negligible effect on the density while having a significant effect on the electrolyte compressibility. For a defined liquid volume of fixed height:

$$f = \frac{f_w}{(1 + f_a \frac{K_a}{K_w})^{1/2}}$$

where $f_w$ is the resonance frequency of the fundamental resonance mode in pure water (also indicated as volume line, because it depends on the liquid level height and $f$ is the resonance frequency in bubble-filled water that is measured directly using BARDS in this work. Here, the compressibility ratio is defined by:

$$\frac{K_a}{K_w} = \frac{v_w^2 \rho_w}{\gamma P}$$

where $K_a$ is the adiabatic compressibility of dry air, $K_w$ is the compressibility of water, $\rho_w$ is the water density, $\gamma$ is the ratio of specific heats for dry air, and $P$ is the atmospheric pressure. In this case, $K_a/K_w = 1.49 \times 10^4$. The value $\gamma$ is 1.40 for dry air. The corresponding values for $H_2$ (1.405) is almost identical leaving the ratio essentially unchanged. The approach then allows direct monitoring of $f_a$, the fractional volume occupied by hydrogen bubbles in a half cell arrangement. The experimental setup allows for the resonance frequency of a given system to be monitored over time. In the context of electrolysis, the technique is used here to compare the volume of gas evolved by different electrodes, which measured the effect of gas escape, trapping in the microlattice, and reabsorption in solution over time. The microlattice electrodes were analyzed with this technique in a two-electrode configuration, with the microlattice acting as the cathode, and a platinum wire acting as the anode. To ensure acoustic response from only hydrogen evolution, the electrodes were separated via a salt-bridge. The electrolyte was pre-stirred to ensure any dissolved gases were removed prior to testing.

*Modelling of Larger Bubble Traffic*

A computational model was developed to simulate gas bubble transport, trapping, and escape within the Octahedra and OctetBig microlattice electrode structures, reconstructed from their STL mesh used for printing. The geometry and the solid lattice boundaries were used to define the available pore space through which bubbles move. To quantify pore-scale constraints on transport, the local pore-throat diameter field $d_t(x_i, y_j)$ was estimated using a vertical ray-casting procedure. For a grid of lateral coordinates $(x_i, y_j)$, upward rays $r(t) = (x_i, y_j, z_0 + t)$ were intersected with $\Omega_s$, producing a set of intersection points:

$$z_k: r(t_k) \in \partial\Omega_s$$

The pore throat at that lateral position was defined as the smallest solid-to-solid spacing:

$$d_t(x_i, y_j) = \min_k (z_{k+1} - z_k),$$

which serves as a local upper bound bubble diameters that are allowed to pass through.

Gas bubbles were represented as spherical particles characterized by position, radius, and velocity. Initial bubble populations were randomly seeded within the lower region of the lattice, and additional bubbles were continuously introduced during the simulation to mimic sustained gas evolution. Bubbles were modeled as non-interacting spheres with position **x$_i$(t)**, radius **R$_i$**, and velocity **v$_i$(t)**. Each bubble's motion followed a kinematic update:

$$\mathbf{x}_i(t + \Delta t) = \mathbf{x}_i(t) + \mathbf{v}_i(t)\Delta t$$

where *Δt* = 1 simulation step. The velocity consists of a prescribed upward drift term **v$_0$** and a stochastic dispersion term ξ(t):

$$\mathbf{v}_i(t) = v_0\,\mathbf{e}_z + \boldsymbol{\xi}(t), \boldsymbol{\xi}(t) \sim U(-\epsilon, \epsilon)^3.$$

Boundary reflections at $\partial\Omega_s$ were defined as:

$$\mathbf{v}_i^{\text{new}} = \mathbf{v}_i^{\text{old}} - 2(\mathbf{v}_i^{\text{old}} \cdot \mathbf{n})\mathbf{n},$$

for any collision normal **n**. Given a lattice dimension of 10 mm, and a vertical velocity defined in the model as 0.002 units per frame (based on average bubble velocity in water for these size bubbles or 100 mm/s), each frame here represents ~ 0.2 ms per frame. For 1000 frames, this represents 0.2 s of actual time for a single bubble at this velocity to reach the upper part of the lattice in a straight line, unimpeded.

Bubbles were immobilized ("trapped") when their diameter exceeded the local pore-throat dimension obtained from the ray-cast map. A bubble becomes trapped when its diameter exceeds the local pore-throat diameter:

$$2R_i \geq d_t(x_i, y_i).$$

Once trapped, its velocity is set to zero

$$\mathbf{v}_i(t) = \mathbf{0},$$

and the bubble remains immobilized at its trapping location. Unlike some coalescence models, bubble merging was explicitly disabled to isolate the effects of geometric confinement within the microlattice structure.

Bubbles exiting the top of the lattice domain were classified as "escaped" and kept in their escaped position to preserve their spatial distribution for visualization and statistical analyses. The escaped bubble is defined when:

$$z_i(t) > z_{\max},$$

where $z_{\max}$ is the upper boundary of $\Omega_p$. Escaped bubbles are retained in the visualization domain with:

$$\mathbf{v}_i(t) = \mathbf{0}, \text{state} = \text{escaped}.$$

The model tracked the full trajectory history of every bubble, enabling the computation of mean squared displacement (MSD), residence time distributions (RTD), and frame-resolved counts of free, trapped, and escaped bubbles. Additionally, spatial trap-density heatmaps, clogging rates ($\Delta$ trapped per frame), and escape rates ($\Delta$ escaped per frame) were obtained. The model records the complete trajectory of every bubble. A frame-resolved mean squared displacement (MSD) was computed as:

$$\text{MSD}(t) = \frac{1}{N(t)} \sum_{i \in \mathcal{A}(t)} \| \mathbf{x}_i(t) - \mathbf{x}_i(t_0) \|^2,$$

where $\mathcal{A}(t)$ is the set of active bubbles (free, escaped, or trapped). Bubble residence time $\tau_i$ was defined as:

$$\tau_i = t_{\text{exit},i} - t_{\text{birth},i},$$

with exit defined as either trapping or escape. Frame-resolved population counts were computed for free, trapped, and escaped bubbles:

$$N_{\text{free}}(t), N_{\text{trapped}}(t), N_{\text{escaped}}(t).$$

Clogging rate was quantified using the discrete derivative:

$$\Delta N_{\text{trapped}}(t) = N_{\text{trapped}}(t) - N_{\text{trapped}}(t-1),$$

and similarly for escape rate.

## Results and Discussion

*3D printed microlattice electrodes and electrochemically active surface areas*

For this work, the printed microlattices acted as the working electrode, which reduces water to produce hydrogen gas and hydroxide ions via the hydrogen evolution reaction (HER). The lattice electrodes have high surface areas, allowing for a higher rate and volume of hydrogen production when compared to a planar electrode of similar dimensions. The 3D graphical rendering of each design can be seen in Figure 1(a). Their geometric structures were chosen specifically to have a defined and controlled porosity to examine the influence of surface area and pore size on bubble nucleation, bubble traffic, detachment, efficiency, stability and gas trapping/reabsorption effects via electrochemistry and broadband acoustic resonance dissolution spectroscopy (BARDS). The effectiveness of the metallized printed lattices as HER electrodes (current response, electrolysis stability, etc.) was investigated via a series of electrochemical analyses in neutral monobasic/dibasic potassium phosphate electrolyte. These electrodes were designed to examine HER process using primarily open cell porous structures. The actual printed structures can be seen in Figure 1(c) and (d). The geometrical surface areas determined directly from their CAD drawings are providing in the Supporting Information, Table S1. The variation in internal pore density, structure, tortuosity and size specifically targets the influence of bubble transport, traffic and trapping, which is dependent on bubble rate, nucleated and full-grown bubble size and density. Those are in turn driven by the activity of the material and the electrochemical parameters (current, voltage). In experiments, the microlattice structures coated in NiO/Ni(OH)$_2$ (referred to as Ni-coated) and Au, are tested for hydrogen evolution in a three-electrode cell (Figure 1(e)).

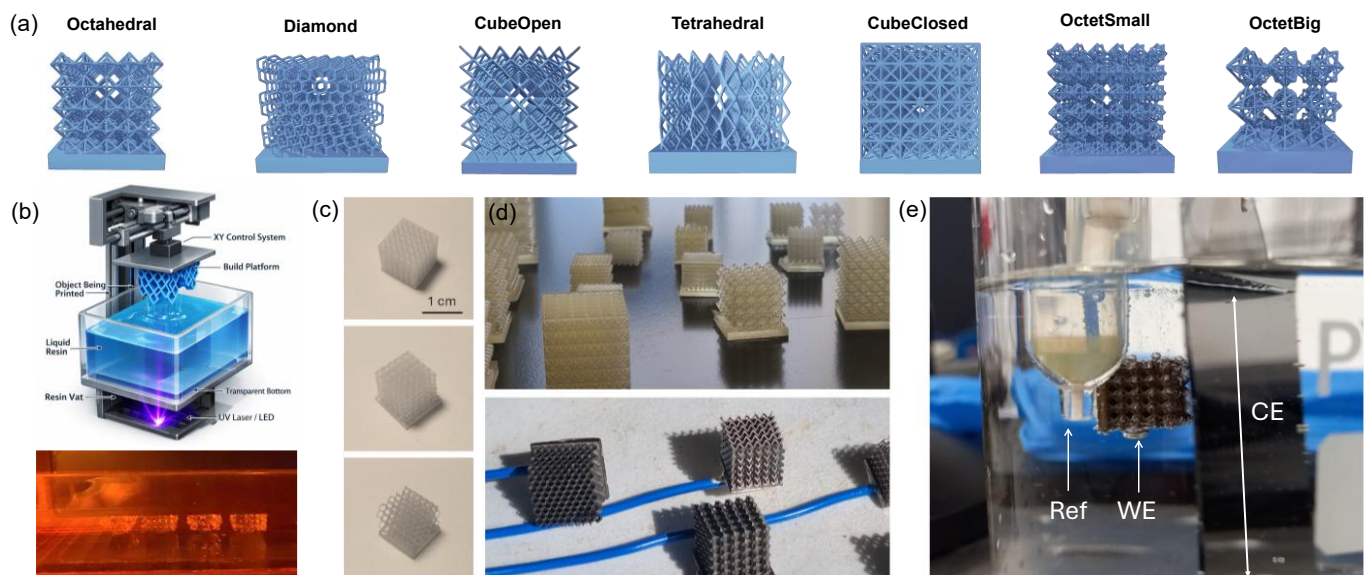

**Figure 1** (a) Seven microlattice structures were designed for use as water-splitting electrodes in this work, specifically to investigate hydrogen evolution. (b) Schematic for a typical vat polymerization 3D printer, where PMMA-based resins are stereolithographically photopolymerized. (c) Microlattice electrodes post-printing, cleaning, and UV curing. (d) A variety of printed and processed microlattice structures, with the lower half showing several Ni-coated lattice electrodes with their attached ohmic contact. All lattices tested in this work have dimensions of approximately 1 × 1 × 1 cm (volume of ~1 cm$^3$). Sputtering is used to uniformly coated the entire surface, which is effective for more open designs. (e) The three-electrode setup used for electrochemical analyses in this work. The metallized printed lattice in the centre acts as the working electrode (WE), the platinum-coated silicon wafer acts as the counter electrode (CE), and the saturated calomel electrode (SCE) acts as the reference electrode.

To determine the electrochemically active surface area (EASA) of our metallized lattice electrodes in the neutral pH electrolyte, a series of non-faradaic CV measurements were carried out at scan rates ranging from 10 to 100 mV/s between -0.241 – -0.041 V (SCE) for Ni-coated microlattices and between 0.009 – 0.209 V (SCE) for Au-coated microlattice electrodes, and with planar electrodes of each material. To establish a well-defined planar reference, identical sputtering conditions were applied to coat a known area of single-crystal Si(100), providing a baseline current density from an atomically flat surface of known geometric area. The non-faradaic CV responses of the Ni- and Au-coated Si(100) wafers are shown in Figure 2(a) and (d), respectively. Corresponding CV curves for the Ni- and Au-coated microlattice electrodes, using the CubeOpen architecture as a representative example, are shown in Figure 2(b) and (e). Cyclic voltammetry was acquired for all structures, and the cumulative charge (CV area) were plotted as a function of scan rate for the Ni-coated (Figure 2(c)) and Au-coated (Figure 2(f)) electrodes. The slopes of these linear relationships were extracted to quantify the electrochemical response and to determine the electrochemically active surface area (EASA) as a ratio relative to the geometric surface area (GSA) of each microlattice. Representative CV curves carried out at 50 mV/s for all microlattices are shown in the Supporting Information, Figure S1. The non-faradaic CV curves in Figure 2 (a,b) and (d,e) are purely an electronic double-layer capacitance (EDLC) response, which is directly proportional to the EASA of a given electrode. As the EASA of both sputtered wafers are known, the ratios of slopes are used to calculate the EASA of all printed microlattice electrodes shown in Figure 1. The slopes, ratios, and EASA values are also tabulated in the Supporting Information, Table S1.

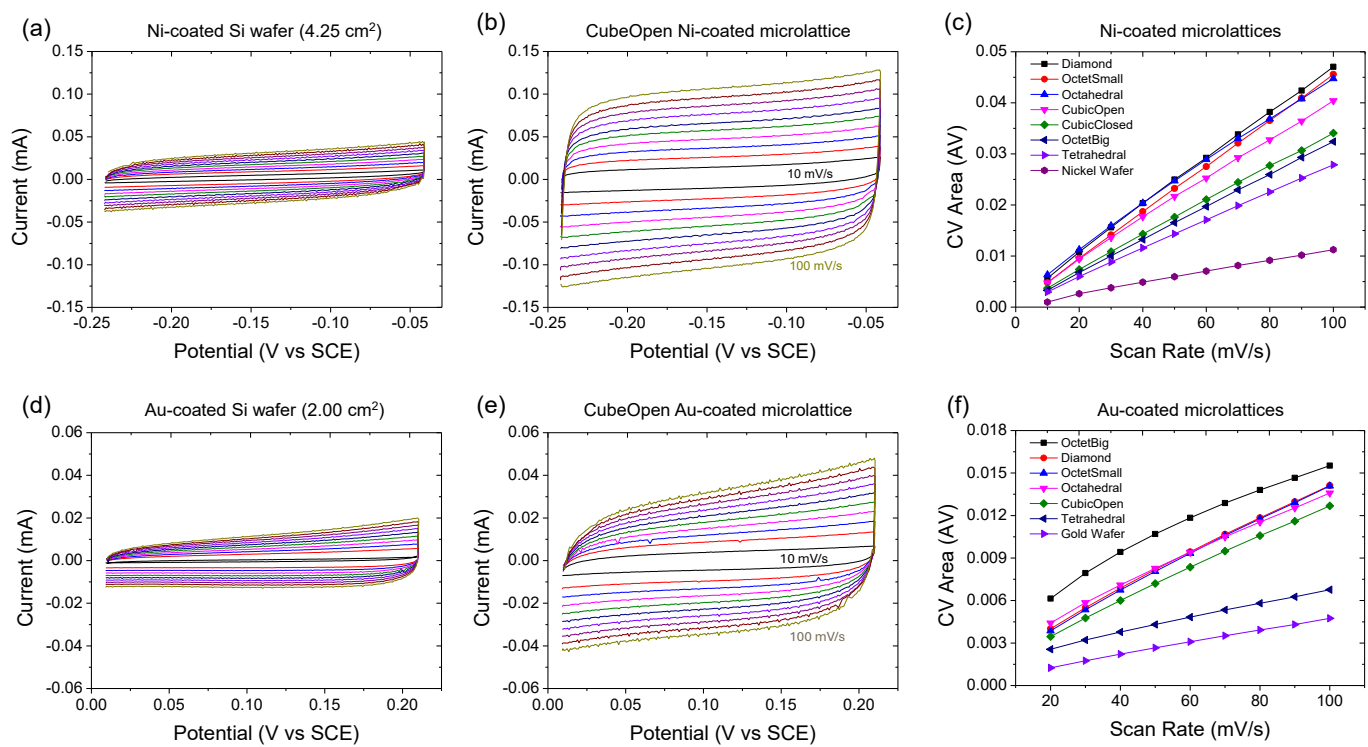

**Figure 2.** Cyclic voltammetry of (a) Ni-coated and (d) Au-coated Si(100) wafers carried out at scan rates of 10–100 mV/s between -0.241 – -0.041 V (SCE) and 0.009 – 0.209 V (SCE), respectively. Cyclic voltammetry curves under identical conditions are also shown for (b) CubeOpen Ni-coated and (e) CubeOpen Au-coated microlattice electrodes. (c) Comparison of cumulative charge vs scan rate for all Ni-coated and (f) Au-coated microlattice electrode structures using identical procedures used in the data presented in (a,b and d,e) used to determine the electrochemically active surface area (EASA) using the response on pristine atomically flat coated Si(100) wafers as a surface area reference.

The calculated EASA values, compared with estimates based on GSA, are presented in Figure 3. Figures 3a and 3b compare the EASA values of Ni-coated and Au-coated microlattice electrodes, respectively, with their corresponding GSA values, which were computed from the digital microlattice designs assuming perfectly smooth surfaces. For Ni-coated structures, four architectures (Diamond, Octahedral, CubicOpen, and OctetBig) exhibit EASA values exceeding their GSA, whereas the remaining three (OctetSmall, CubicClosed, and Tetrahedral) show lower EASA than predicted geometrically. In the case of the less electrochemically active Au-coated microlattice, all structures exhibit effective EASA values from voltammetric measurements lower than the corresponding GSA values. To understand the differences, we first investigated the morphology of the printed microlattice trusses/struts and the deposited coatings. Figures 3(c) and (d) show a series of printed surface features (pores, bumps, ridges, etc.) of Ni-coated microlattices. In Figure 3(e), we see a section of Au-coated trusses where the microporosity can be seen in cross section. These ridges and pores increase the effective area of the microlattice structures when coated, compared to the GSA values (which assume ideally smooth and geometrically perfect architectures). The inset to Figure 3(a) shows a cross-section of the Ni-coated CubeClosed microlattice electrode. This enclosed-type structure causes obvious issues for complete metallization, as the sputtering process is unable to penetrate deep into the lattice structure. This resulted in incomplete metallization of the lattice structure. Of the 7 microlattice structures, this is the only structure where incomplete metallization occurs, but we note that fully metallized structures such as the OctetSmall and Tetrehedral structures also show suppressed EASA value compared to the GSA response, and thus incomplete coverage does not account for reduced overall HER gas volume/activity alone.

Under identical conditions, the structured porous microlattices have a specific influence on bubble generation and trapping that influence the overall electrochemically active surface area. As the surface of the deposited coatings is greater in roughness compared to the planar Si(100), we posit that the inherent HER activity is affected primarily by bubble coverage of the active material in some structures. This effectively reduces the EASA below the values of the GSA in the case of Ni-coated microlattices in two of the geometries; the third most closed, small-cell microlattice is compromised by incomplete internal coating. For the other four microlattices, the HER bubble traffic is comparatively unimpeded, and the roughness of the deposited coatings and electrochemical activity contribute to the enhancement of the total charge using the microlattice geometry. Using porosity on two distinct length scales (in the trusses and in the overall microlattice architecture) the EASA is improved beyond the surface area of the microlattice, with minimal bubble trapping. This approach allows for conformal coating of HER materials, with control of porosity and its function dictated by additive manufacturing of the substrate instead of micro/nano-structuring the active material itself. When the material is significantly less electrochemically active, such as Au in this case, the benefit of the smaller length scale roughness on gas evolution during HER is limited. For Au, under identical electrochemical conditions, the rate generation of hydrogen bubbles is much less. Bubbles are fewer, nucleated more slowly and detach from the surface as larger bubbles. In all microlattices investigated here, we see lower EASA values compared to the GSA values when the electrochemical activity is lower (such as Au in this case), increased bubble size and trapping that reduces active surface area overall.

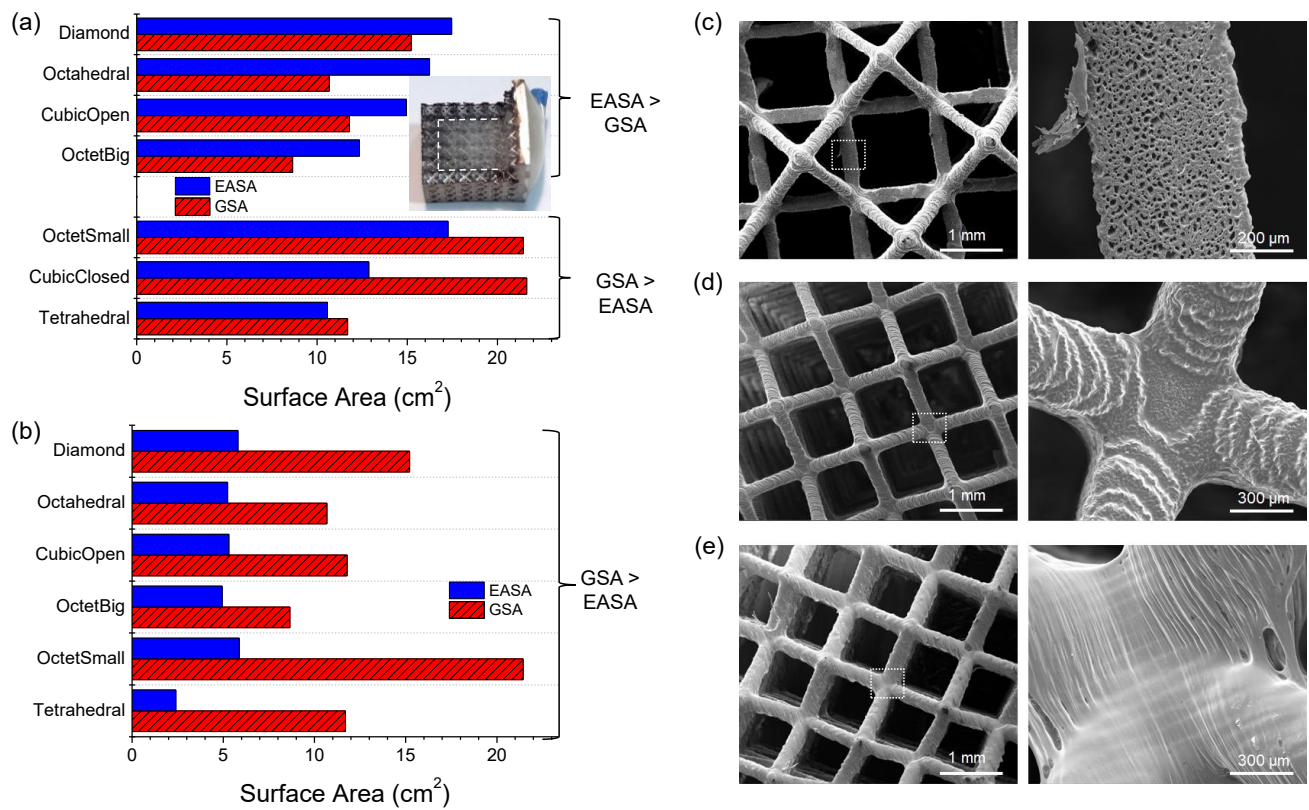

**Figure 3.** Comparison of the electrochemically active surface areas (EASA) and geometric surface areas (GSA) for (a) all the Ni-coated and (b) Au-coated microlattice electrodes. (Inset) Cross-section of a Ni-coated CubicClosed lattice electrode, where the more closed porous structure prevented internal coating. SEM images show printing imperfections in both the (c,d) Ni-coated and (e) Au-coated lattice electrodes, which contribute to the EASA.

*Electrochemical response of Ni- and Au-coated microlattices*

We obtained linear sweep voltammograms for all microlattices to examine HER reactions. Figure 4(a) shows the response of all Ni-coated microlattice electrodes, where the onset potential is similar across all microlattice architectures (between -0.44 and -0.48 V (RHE)) as shown in Figure 4(b). In many high surface area and porous HER materials, the influence of pore size, distribution and surface area is critical for reducing the onset potential for efficient hydrogen evolution. These microlattices are designed with a volume that examines large bubble traffic and trapping. When the response is also shown as a function of current density in Figures 4(a) and (b), this influence can be seen. Due to unimpeded bubble traffic and detachment governed by several existing models (*43*), the Ni-coated Si(100) wafer electrode shows a higher maximum current density (-5.63 mA/cm$^2$ at -0.8 V (RHE)) and earliest onset potential for a given current density of -1 mA/cm$^2$ at -0.60 V (RHE). Of the microlattice electrodes, the OctetBig achieves a current density of -4.12 mA/cm$^2$ at -0.8 V, and -1 mA/cm$^2$ at -0.63 V. With an EASA that is ~12× greater than the Ni-coated wafer, it is clear that bubble traffic, impedance to nucleation, and larger bubble residence times affect what is otherwise a greater electrochemically active surface area under non-Faradaic conditions. During HER, however, even large ordered porous structures (with smaller porosity within the trusses to increase EASA) does not guarantee improved overall gas bubble evolution as typically indicated by current density. Bubble trapping effectively reduces accessible EASA.

To determine the stability of these Ni-coated lattice electrodes in neutral aqueous electrolyte, a CV stability test was carried out following the initial HER LSV shown in Figure 4(a). For the stability test, electrodes were cycled between 0.0 V and -0.8 V (RHE) 100 times at a scan rate of 100 mV/s, after which another HER LSV was carried out (shown in Figure 4(c)), with identical parameters to the initial HER LSV test. For all microelectrodes, repeated polarization between 0 V and the lower voltage limit of -0.8 V, degraded the coatings, such that the OctetBig and the flat Ni-coated Si wafer retained only 28% and 33% (respectively) of their initial HER current density at the same voltage. The effect of large bubble trapping on EASA is not linked to coating degradation, since fresh electrodes were always used when assessing the response as a function of EASA and GSA.

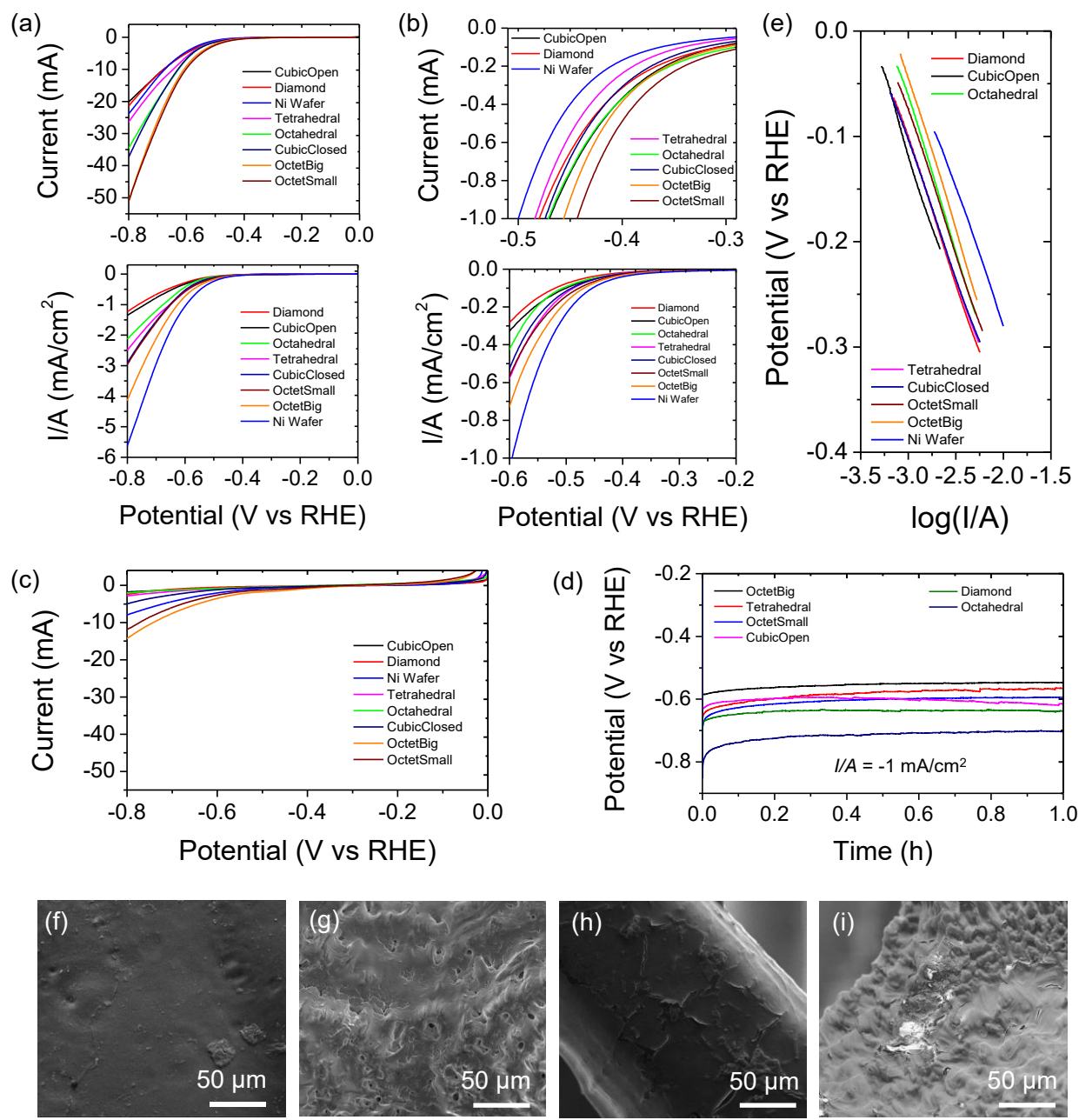

**Figure 4** (a) Linear sweep voltammetry (LSV) responses of all Ni-coated microlattice electrodes, and (b) onset potential region. The same data is also presented underneath as current density factored using the microlattice EASA. Electrodes were swept at a scan rate of 5 mV/s from a potential of 0.0 V to -0.8 V (RHE). (c) Overall LSV current responses of all Ni-coated microlattice after being subjected to a CV stability test. For the stability test, electrodes were cycled between 0.0 V and -0.8 V (RHE) 100 times at a scan rate of 100 mV/s. Immediately after, a post-stability LSV was carried out under identical conditions to the initial LSV test in (a). (d) Chronopotentiometry of all Ni-coated microlattice electrodes. A constant current density of -1 mA/cm$^2$ was applied to all the electrodes for a duration of 1 h. (e) Tafel plots for all microlattice electrodes from the data in (a). The current density was factored using the EASA of each electrode. SEM images show the nickel coatings of different electrodes: (f) before any electrochemical tests, (g) after the initial LSV test, (h) after the stability CV test, and (i) after the chronopotentiometry test. Image (h) shows some initial signs of degradation and peeling of the Ni coating, resulting in the reduced current responses in the post-stability LSV tests shown in (c). The coating in image (i) shows extensive cracking and peeling of the nickel layer, with exposed polymer underneath the Ni-coating layer.

Chronopotentiometry tests were carried out on fresh nickel-sputtered electrodes. A constant current density of -1 mA/cm$^2$ was applied to the electrodes for 1 h. The potential versus time responses for these tests are shown in Figure 4 (d). As with the previous Ni-coated microlattice electrochemical tests, the OctetBig-designed microlattice requires the lowest overpotential to achieve a constant current density of -1 mA/cm$^2$. All microlattice electrodes are shown to be stable during the constant electrolysis of this test. To examine the kinetics of the HER process and gauge the influence of the microlattice on overall bubble evolution, traffic and efficiency in a neutral pH electrolyte, we show Tafel plots of relevant potential regions in Figure 4(e). Following best-practices for Tafel analysis (*44-47*), the data shown Tafel slopes for all microlattice electrodes in the range 250 – 280 mV/dec, with transfer coefficients ranging from 0.21 – 0.24. This high Tafel slope is indicative of sluggish kinetics and indicates that active materials with localized high surface area evolve hydrogen that is subsequently limited by bubble size and long residence times typical of transport limitations through the tortuous microlattice architecture. It is notable that the lower of the Tafel slopes are from the Octahedral and OctetBig microlattices, which provides the least impedance to bubble traffic. However, the lattices fundamentally influence overall bubble release from HER even for active materials with relatively high EASA, to an extent that the overall response is essentially independent of geometric and electrochemically active surface areas, instead dominated by diffusion-limited transport sluggishness of bubble transport out of the microlattices. This is in marked contrast to some reports, where ordered porosity in 3D lattices promotes efficiency bubble transport. However, bubble size effects and reduced tortuosity in vertically aligned pores are likely to be influential for bubbled formed in strongly alkaline electrolytes in that case (*30*).

To further examine the effect of different electrochemical conditions on the Ni-coated electrodes, SEM was used to obtain images of a variety of nickel-based coatings, shown in Figure 4 (f)–(i). As the images were taken across several microlattices, morphological differences (pores, bumps, etc) are due to printing, rather than electrochemical processes. Both the freshly coated (Figure 4(f)) and post-HER LSV (Figure 4 (g)) coatings are pristine, with no signs of cracking, peeling, or degradation. The post-stability CV Ni-based coating (Figure 4(h)) shows initial signs of degradation, with small amounts of cracking and exposed polymer visible. The post-chronopotentiometry Ni coating (Figure 4(i)) has suffered the worst degradation, with cracking and peeling visible across the surface, exposing the most polymer underneath.

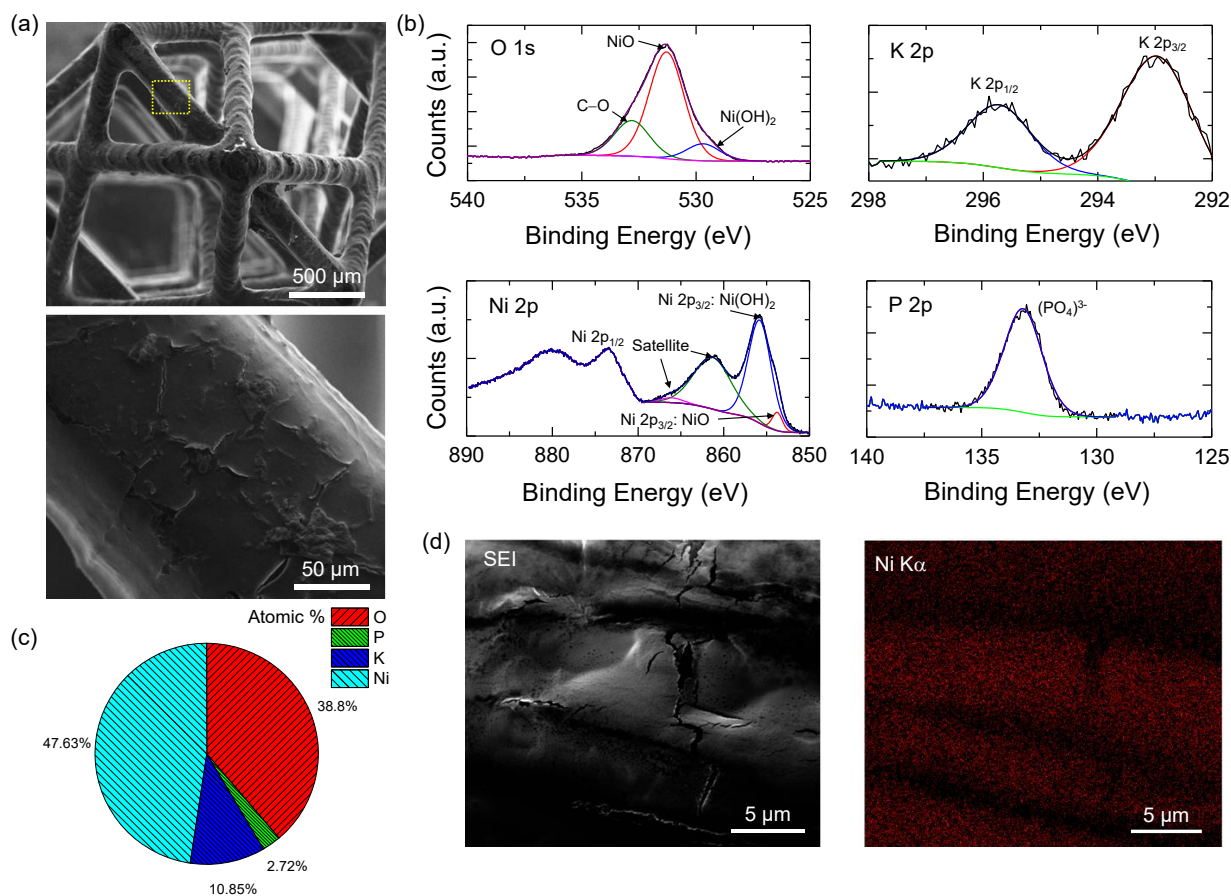

**Figure 5.** (a) SEM images of a Ni-coated microlattice electrode after the stability CV test. (b) Photoemission spectra from X-ray photoelectron spectroscopy measurements of O 1s, K 2p, Ni 2p and P 2p core-levels. (c) Atomic % EDX analysis for the same Ni-coated microlattice with corresponding (d) secondary electron image and mapping.

We examined the surface composition in more detail to determine the active Ni phase that results in HER processes. In Figure 5(a), we show SEM data of a microlattice electrode after being subjected to the stability test outlined in Figure 4(c) whereby the electrodes underwent a CV between 0.0 V and -0.8 V (RHE) 100 times at a scan rate of 100 mV/s, and then an LSV from 0.0 – -0.8 V. Some cracking is evident, but no delamination has occurred after extended HER bubble evolution. XPS analysis shown in Figure 5(b) confirm the Ni-coatings contain NiO and Ni(OH)$_2$. It is notable that the Ni 2p core level photoemission shows no evidence of metallic Ni$^0$. From the Ni 2p core-levels, we see a higher intensity of the Ni(OH)$_2$ photoemission signal from the hyperfine split 2p doublet, indicating a near surface presence of Ni(OH)$_2$. NiO is also detected and we posit that it is buried as a subsurface layer under the Ni(OH)$_2$ given its lower intensity in the Ni 2p spectrum (*48*). The O 1s core level spectrum, which is sensitive to the oxygen binding environment, shows a dominant intensity for the NiO phase, which is consistent with bulk-like lattice oxygen. A lower intensity contribution is found for Ni(OH)$_2$ compared to Ni 2p from the same measurement, despite the species being closer to the surface as evidence by the Ni-sensitive 2p spectral data. XPS measurements also detected potassium and phosphorus, as expected from remnant dried salts from the electrolyte, and core-level emission consistent with a phosphate phase. Corresponding C 1s core levels are provided in the Supporting Information Fig. S2. Energy dispersive X-ray analysis in Figure 5(c) with secondary electron images close to a crack in the film (Figure 5(d)) are consistent with the Ni(OH)$_2$-NiO coating layer.

Following the analyses for the microlattice electrodes with Ni-based coating (Figure 4), an identical series of electrochemical tests were conducted on Au-coated microlattices. The activity of pure Au (sputtered film, not nanoscale with active sites, nor alloyed) is considerable weaker than NiO or Ni(OH)$_2$ for HER in neutral pH electrolytes. These measurements addressed the relationship between intrinsic activity of the coating and the extrinsic EASA and geometric surface area and architecture influence on bubble nucleation, evolution, and traffic with respect to more active materials on identical microlattice under identical conditions. Figures 6(a) and (b) show the response of all Au-coated microlattice electrodes after undergoing an LSV from 0.0 – -0.8 V. We note that the current measured is ~2.5× lower compared to Ni-coated lattices under identical conditions, and this is attributed to the lesser activity of the Au. However, when the EASA is measured separately from these microlattice electrodes and factored, the current density from all lattices has a very similar value to the Ni-coated microlattice electrodes in Figure 4(a) and (b). The Octahedral microlattice was found to be the most HER active with an Au-coating after its first LSV sweep. And as shown in Figure 6(b), this microlattice also showed the lower overpotential for onset of HER, achieving a current response of -1 mA at -0.61 V vs RHE. The OctetBig and Tetrahedral microlattices, as open structured architectures, also provide the higher current densities when coated by Au. The difference in intrinsic HER activity is found by comparing the response of flat films of both metals, where we see an Au-coated Si(100) wafer shows the lowest current and current density values, opposite to that of Ni-coated flat films seen in Figure 4. Thus, less active HER materials are understandably improved by increasing geometrical surface area using microlattices in this case. For Au, under identical electrochemical conditions, the rate generation of hydrogen bubbles, and their quantity, is much less. Their nucleation and detachment from the surface is also slower compared to the Ni case, where (see Supporting Information Movie S1) a clear effervescence occurs. For Au, this process is extremely slow, showing fewer, larger bubbles. The present work shows that the Octahedral architecture seems to favour the traffic of larger bubbles under identical conditions, with lower residence times or trapping that could lead to reduced active area for HER. As we saw for Ni-coated structures, a slightly difference architecture (OctetBig) was needed to maximize activity overall.

Chronopotentiometric tests were also carried out, with a constant current density of -1 mA/cm$^2$ applied to the Au-coated microlattice electrodes for an hour. The potential versus time responses for these tests are shown in Figure 6(c). As with the LSV in Figure 6(a) and (b), we find that the Octahedral designed microlattice requires the least negative overpotential to achieve a constant current density of -1 mA/cm$^2$. All other lattice electrodes are shown to be stable during the constant electrolysis of this test. SEM was carried out to examine the gold coatings after electrochemical testing for HER, shown in Figure 6 (d)–(g). As the images were taken across several lattices, morphological differences (pores, bumps, etc) are due to printed morphology, and not because of electrochemical processes such as corrosion, etching, delamination or pitting. Unlike the Ni-based coatings, which showed clear signs of degradation after the chronopotentiometry and stability CV tests, no such damage to the lower activity Au coatings is found.

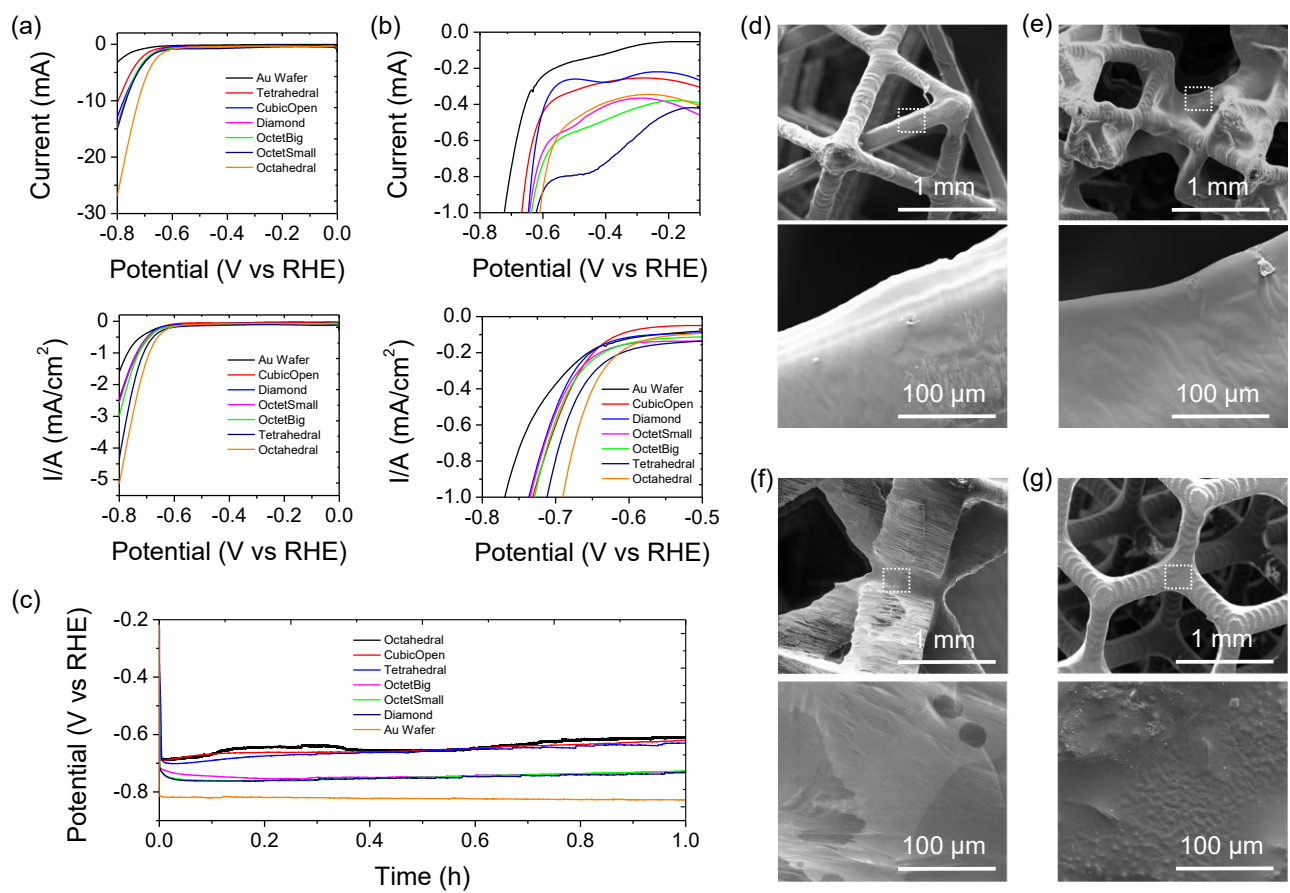

**Figure 6.** (a) Linear sweep voltammetry (LSV) responses of Au-coated microlattice electrodes, and (b) onset potential region. The CubicClosed lattice was not used here since the lattice cannot be fully coated throughout. The same data is also presented underneath factored using the microlattice EASA. Electrodes were swept at a scan rate of 5 mV/s from a potential of 0.0 V to -0.8 V (RHE). (c) Chronopotentiometry of all Au-coated microlattice electrodes. A constant current density of -1 mA/cm$^2$ was applied to all the electrodes for a duration of 1 h. SEM images show the Au coatings of different electrodes: (d) before any electrochemical tests, (e) after the initial LSV from 0.0 to -0.8 V (RHE), (f) after the chronopotentiometry measurements in (c), and (g) after the stability CV test where the microlattice was subjected to an LSV to -0.8 V after 100 CV cycles between 0.0 V and -0.8 V (RHE) at a scan rate of 100 mV/s.

*Influence of microlattice architecture on bubble transport*

We also modelled gas traffic through the microlattice architectures. Based on the response of Ni-coated electrodes, we chose the OctetBig and the Octahedral structures to examine bubble traffic in the best microlattice designs shown in this work. It is reasonable to assume a relatively consistent microporosity and roughness on the trusses for the Ni-coated microlattice surface in each design (compared to the length scale of porosity of the architectures themselves). We examined the influence of the pore size, structure, orientation in 3D space, the tortuosity and effect on bubble residence times for various bubble sizes by modelling the parameters outlined in detail in the Experimental section. Bubble sizes were varied between 0.05× and 0.5× of the subunit that defines each lattice. This corresponds to sizes in the range ~10 – 150 μm. These were determined based on existing reports of bubble nucleation and growth models in the literature (*11, 49-51*). We also examined images from still frames taken of these microlattices undergoing HER process, which shows a typical fully developed bubble size of 100 – 500 μm (*52*), and so the range of sizes proposed above covers bubbles formed at a range of applied currents. We chose to ignore the many small bubbles that easily diffuse through the lattice, and instead concentrated on larger bubble nucleated at that size, or formed due to coalescence, since they are more likely to influence overall bubble traffic and electrolyte-accessible active surface area (*53*).

In Figure 7(a), we show the initial and final states of the modelling with bubbles chosen across the size range at random, nucleated in random positions. Bubbles are made to continually form over the analysis timeframe of 0.2 s, which corresponds to the approximate time it takes an unimpeded bubble to travel a maximum of 1 mm vertical distance and from a starting seed of 50 bubbles, the simulation runs until 500 bubbles are formed. Bubble trapping is evident in both microlattices, with the Octahedral microlattice exhibit a lower mean square displacement over time, shown in Figure 7(b), than the OctetBig, corresponding to a greater number of trapped bubbles, and fewer escaped bubbles than its OctetBig counterpart (Figure 7(c)) whose subunits are larger and the pore throat sizes are greater. While the differences appear small, the microlattice pore structure clearly influences large bubble escape rate. Indeed, clogging rates shown in Figure 7(d) for these continuously nucleated larger bubbles show that trapping and overall bubble traffic occurs very soon, before any appreciable large bubble escape occurs. It further occurs throughout the microlattice and the escape rates in Figure 7(d) confirm that bubble escape is more uniform near the upper surface. Of the 500 bubbles formed across the larger diameter ranges in both microlattices, ~24-26% are predicted to escape, and this value remains consistent for the same parameters over longer durations. Thus, the microlattice structure has a definitive influence on the traffic of larger bubbles, and the degree of bubble trapping is consistent with the experimental data shown earlier. Indeed, for other microlattice structures, where the exit throat of the porous structure is smaller, bubble escape is more difficult/limited and the transition from trapping to detachment and wicking is limited or prevented (*49, 54*). However, smaller bubbles from HER 'effervescence' still occurs.

In practical terms, HER bubble evolution is continuous and while a range of bubble sizes occurs, process that lead to large bubbles clearly do not completely suppress further bubble nucleation, growth, transport and detachment over time. The heat map of the system shown in Figure 7(e) shows that bubble trapping is greatest at locations close to intersection truss nodes throughout, and do not show a preference for increased bubble traffic near the top-most surface where, in this model, bubbles eventually escape. While pore size and other attributes may be nominally important, bubble growth that continues within an open volume may eventually be constrained from detachment by the geometry of the surrounding lattice design.

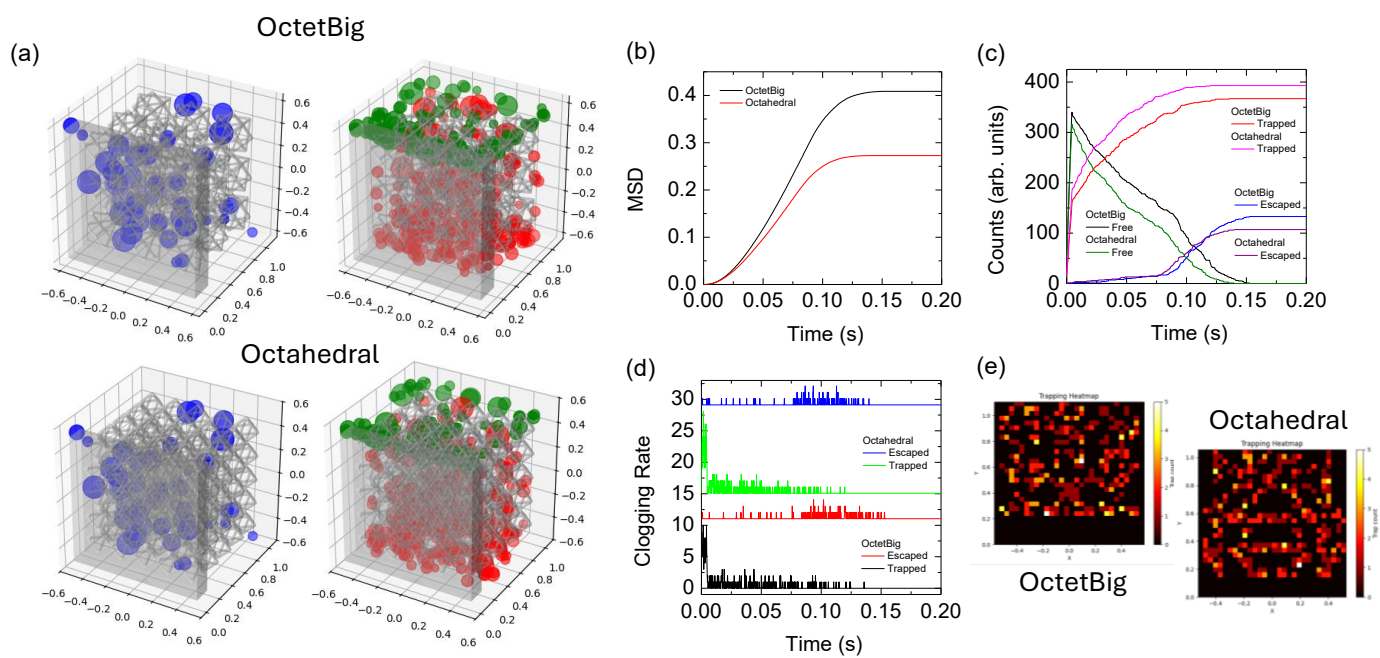

**Figure 7.** (a) 3D render of initial large bubble nucleation with the OctetBig and Octahedral microlattices. (b) Mean square displacement calculations for all bubbles over the duration of the simulation. (c) Populations of simulated trapped, free (not trapped nor escaped), and escaped bubbles from each microlattice. (d) Clogging rates (trapping) and escape rates of all 500 bubbles over the duration of the simulation and (e) heat maps of spatially resolved bubble trapping locations within each microlattice.

Finally, HER processes, bubble generation and the influence of the microlattice on bubble traffic/trapping was analyzed in operando, using BARDS. This approach allows direct measurement of total evolved gas volume, and factors all the gas evolved during HER for all bubbles sizes. This includes nucleated bubbles that form liberated (capturable) gas, and also those bubbles that constitute redissolved gas. Details on this method and the background theory can be found elsewhere (*42, 55*). Accurate electrochemical assessment using established benchmarks and procedures (*45*) is often very useful for many materials and systems, but purely electrochemical methods do not directly measure all gas (including redissolved gas), nor track the real-time effect of bubble density, bubble size, residency within surface features or pores, and other effects. For these operando acoustic resonance measurements, we use the more HER-active Ni-metal coated microlattice electrodes, that we demonstrated earlier to have specific dependence on the geometry of the microlattice porous architecture.

Figure 8(a) shows a series of BARDS spectra for each of the microlattices (Ni-coated) obtained during HER. It is immediately clear that the Tetrahedral, Octahedral and OctetBig microlattice structures show a characteristic response once the applied current is switched on, and which the system returns to a steady state once the current is switched off. The change in the resonant frequency of sound in the electrolyte that defines this response, is definitively related to bubble formation within the electrolyte and the microlattice. The frequency decreases with an increase in compressibility of the electrolyte due to a corresponding increase $H_2$ volume, and eventually returns to steady state after the current is switched off and new bubble generation is stopped. This steady state represents an equilibrium between gas evolution at the electrodes in the electrolyte and gas leaving the electrolyte at the surface. While it is typical to have HER systems benchmarked in a solution saturated with $H_2$ to ensure adherence to a Nernstian behavior, BARDS is sensitive to residual gas in an electrolyte that affects the compressibility changes necessary to change the acoustic resonance frequency during HER; all electrolytes were shaken prior to measurement to minimize residual gas content, so that any bubble that form are those from HER processes.

Owing to significant larger bubble trapping, reduced active surface area, the first three microlattices (Diamond, CubeOpen and OctetSmall) show negligible response. This does not indicate that HER is not occurring (bubbles can be seen during the measurements), but that the quantity of bubbles being evolved, detached and relieved form the microlattice to the electrolyte surface is significantly suppressed compared to the other structures, and thus no change in the resonant frequency is easily detectable.

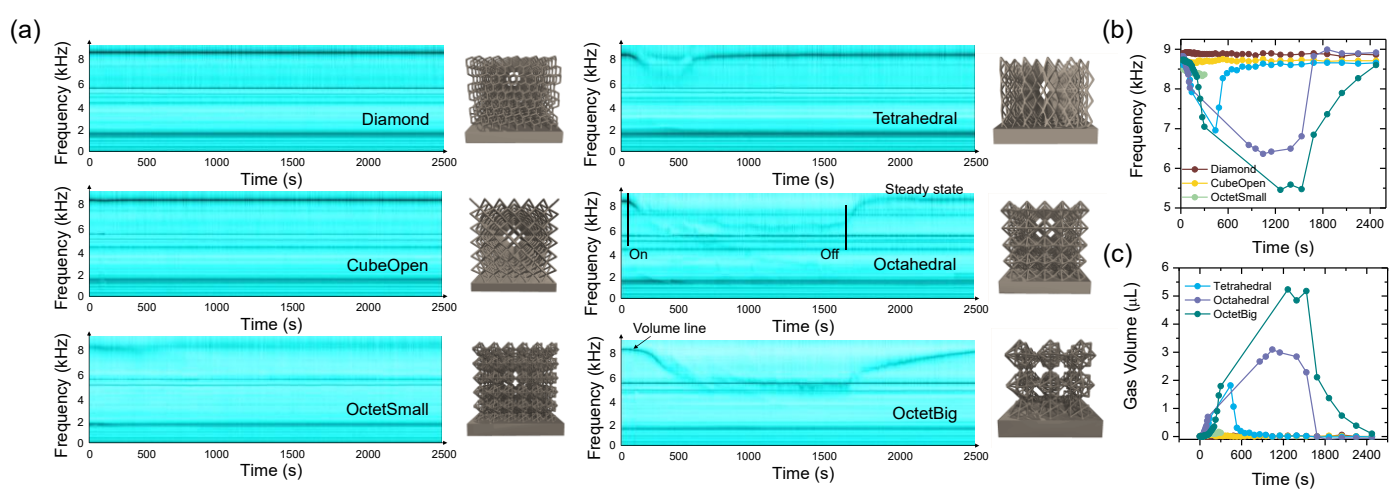

**Figure 8.** (a) BARDS frequency response for each Ni-coated microlattice electrode. Electrodes underwent chronopotentiometry to maintain a constant current density of -1 mA/cm$^2$. (b) Extracted frequency response data for each microlattice and (c) the corresponding gas volume measured using BARDS.

The Octahedral and OctetBig microlattice produced gas bubbles for the longest time (>1500s) compared to all other microlattices. To maintain a pre-defined current density of -1 mA/cm$^2$ in these BARDS measurements (consistent with electrochemical measurements shown earlier), the Off state is determined when a predefined voltage limit is reached, caused by an increase in internal resistance due to bubble coverage and trapping. Longer bubble evolution times for these electrodes indicate a better bubble escape overall. However, as experimental measurements and simulations show, the microlattice structures themselves set limits on the efficacy of bubble transport, detachment and escape to the electrolyte-air interface. By converting the extracted frequency response in Figure 8(b) to evolved gas volume in Figure 8(c), we find a peak gas volume of ~5 μL from the OctetBig microlattice over the 6-minute measurement. The return to a steady state indicates good detachment of bubbles and establishment of overall equilibrium between evolved and dissolved gas. These operando measurements confirm that bubble traffic never completely stalls gas evolution. While active EASA is suppressed by bubble coverage, bubble detachment is the critical factor that relieves active surface area for continued gas evolution.

**Conclusions**

Making porous architectures by additive manufacturing for gas-evolving electrochemical processes such as HER has the potential to improve overall efficiency and capturable gas volume when engineered to optimize both gas evolution *and* detachment. However, they are shown here to have fundamental limitations regarding larger bubbles, or those that form by coalescence of smaller bubbles. By comparing active and less active materials for electrochemical hydrogen evolution in neutral pH electrolyte, we find that electrochemically active surface area is dominated by small scale roughness, while the larger, ordered porosity brings intrinsic problems of bubble traffic through tortuous pore networks and issues with bubble trapping for the larger bubbles. Our data shows that directly sputtered metals and metal oxides such as Au and NiO/Ni(OH)$_2$ are reasonably robust on PMMA-based photopolymerized microlattices formed by vat polymerization. The roughness of the underlying printing process for a given overall geometric surface area of microlattice promotes HER with EASA values well exceeding those from geometrical considerations of the microlattice alone.

All electrochemical measurements show differences due to bubble size, trapping, reduced detachment and build up such that the electrochemically active surface area is compromised to some extent with a large population of large bubbles, which we confirmed by using less electroactive metals that rely on surface area from the microlattice geometry. Modeling simulations also confirm that the larger bubbles are the primary issue, and bubble traffic is largely limited by trapping without sufficient detachment for the larger bubbles until later in time. By then, subsequent gas evolution is limited because the detachment or departure of bubbles is constrained. Operando BARDS data also confirm the underlying issue in microlattice periodic geometries for HER but do show that some structures are better by virtue of their pore structure.

This detailed examination of microlattice electrodes shows that increased surface area and open-cell porous substrates are only one factor in HER electrode design. Careful design is needed to ensure efficient bubble transport, improve detachment and wicking of bubbles, and maximize the use of small-scale surface area to ensure smaller bubble diameters at useful current densities. However, gas evolution in different electrolytes, such as those with innately different viscosity, wettability and other factors may allow microlattice designs to improve the response of new active materials compared to their coating of planar surfaces, especially in scenarios where volume or electrode dimensional constraints are important in various electrochemical apparatus of systems.

**Acknowledgements**

This project has received funding from the European Union Horizon Europe framework under grant agreement no. 101084261 (FreeHydroCells). This publication has also emanated from research conducted with the financial support of Taighde Éireann - Research Ireland Frontiers for the Future Programme under Grant number 24/FFP-P/12834. The support from the Taighde Éireann - Research Ireland Research Centre AMBER (12/RC/2278_P2) is acknowledged. We also acknowledge support from the Irish Research Council under an Advanced Laureate Award (IRCLA/19/118). This work is partly supported by an Enterprise Ireland Commercialisation Fund as part of the European Regional Development Fund under contract no. CF-2018-0839-P. We also thank V. Egorov for assistance in the designs of porous architectures.

# Supporting Information for

# Examination of Hydrogen Evolution Bubble Trapping in Ordered Porous 3D Printed Metal and Metal Oxide-Coated Microlattice Electrodes

Matthew Ferguson[a], Alex Lonergan[a], Christopher Kent[a], Dara Fitzpatrick[a], Colm O'Dwyer[a,b,c]*

[a] School of Chemistry, University College Cork, Cork, T12 YN60, Ireland
[b] Sustainability Institute, University College Cork, Lee Road, Cork, T23 XE10, Ireland
[c] AMBER@CRANN, Trinity College Dublin, Dublin 2, Ireland

*Electrochemical Analyses – Settings*

<u>EASA CV (Ni-coated Microlattices)</u>: microlattice working electrode was cycled within a voltage window of -0.241 V to -0.041 V vs SCE (0.000 V to 0.200 V vs RHE) at scan rates of 10, 20, 30, 40, 50, 60, 70, 80, 90, and 100 mV/s for 5 cycles per scan rate.

<u>EASA CV (Au-coated Microlattices)</u>: lattice working electrode was cycled within a voltage window of 0.009 V to 0.209 V vs SCE (0.250 V to 0.450 V vs RHE) at scan rates of 10, 20, 30, 40, 50, 60, 70, 80, 90, and 100 mV/s for 5 cycles per scan rate.

<u>LSV</u>: lattice working electrode was scanned from -0.241 V to -1.041 V vs SCE (0.000 V to -0.800 V vs RHE) at a scan rate of 5 mV/s.

<u>Stability CV</u>: lattice working electrode was cycled within a voltage window of -0.241 V to -1.041 V vs SCE (0.000 V to -0.800 V vs RHE) at a scan rate of 100 mV/s for 100 cycles.

<u>Chronopotentiometry</u>: a constant current density of -1 mA/cm$^2$ was applied to the lattice working electrode for 1 hour.

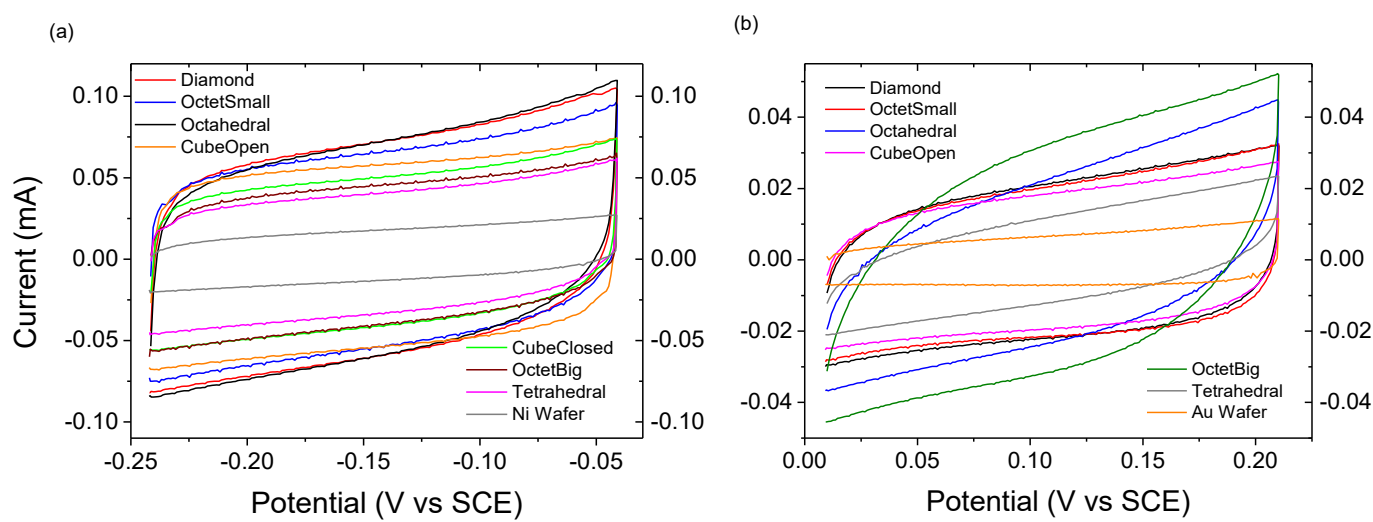

**Figure S1.** EASA CV curves for all (a) Ni-coated and (b) Au-coated microlattice electrodes at a scan rate of 50 mV/s.

**Table S1.** The slopes of the CV area vs. scan rate plots shown in Figure 2 (c) and (f) of the main text are shown below. The ratios of the slopes are calculated, with all Ni-coated microlattices being compared to the Ni-coated wafer, and all Au-coated microlattices compared to the Au-coated wafer. As the EASA values of both wafers are known*, the ratios can be used to calculate EASA values of the lattice electrodes. The GSA values for each printed microlattice geometry design is also shown.

| Coating | Electrode | Slope of CV Area vs. Scan Rate | Ratio of Slope | EASA (cm$^2$) | GSA (cm$^2$) |
|---|---|---|---|---|---|
| Nickel | Nickel-Sputtered Wafer | 0.000110873 | 1.000 | *4.250 | - |
| | CubeOpen Lattice | 0.000390123 | 3.519 | 14.954 | 11.790 |
| | Octahedral Lattice | 0.000423639 | 3.821 | 16.239 | 10.680 |
| | Diamond Lattice | 0.000455495 | 4.108 | 17.460 | 15.220 |
| | Tetrahedral Lattice | 0.000275953 | 2.489 | 10.578 | 11.700 |
| | OctetSmall Lattice | 0.000450428 | 4.063 | 17.266 | 21.44 |
| | OctetBig Lattice | 0.000322142 | 2.906 | 12.349 | 8.650 |
| Gold | Gold-Sputtered Wafer | 0.000043238 | 1.000 | *2.000 | - |
| | CubeOpen Lattice | 0.000114754 | 2.654 | 5.308 | 11.790 |
| | Octahedral Lattice | 0.000113172 | 2.617 | 5.235 | 10.680 |
| | Diamond Lattice | 0.000125537 | 2.903 | 5.807 | 15.220 |
| | Tetrahedral Lattice | 0.000051794 | 1.198 | 2.396 | 11.700 |
| | OctetSmall Lattice | 0.000126935 | 2.936 | 5.871 | 21.44 |
| | OctetBig Lattice | 0.000106678 | 2.467 | 4.934 | 8.650 |

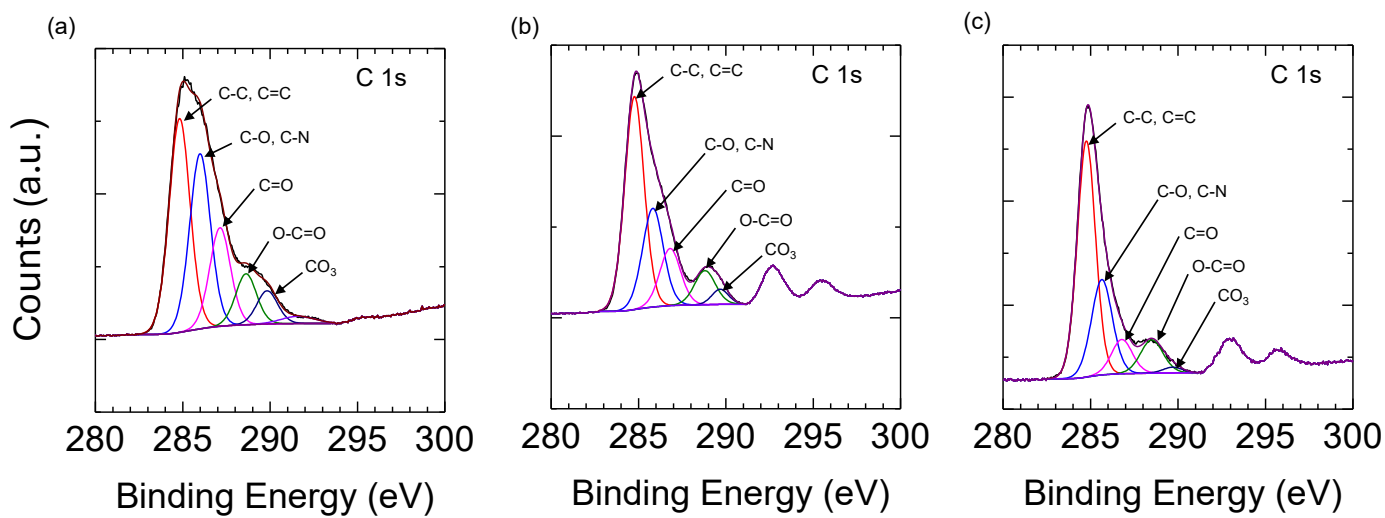

**Figure S2.** Core-level photoemission spectra from X-ray photoelectron spectroscopy measurements of C 1s obtained from Ni-coated microlattices (a) after resting in electrolyte for 1 hour, (b) after resting in electrolyte for 24 hours, and (c) after the stability CV test.